\documentclass[twocolumn,prd,aps,superscriptaddress,preprintnumbers,tightenlines,showpacs,nofootinbib,eqsecnum,amsfonts,amsmath]{revtex4-1}
\usepackage[normalem]{ulem}
\usepackage{placeins}
\usepackage{color}
\usepackage{calc}
\usepackage{amsmath,amssymb,graphicx}
\usepackage{bm}
\usepackage{times}
\usepackage[varg]{txfonts}
\usepackage[colorlinks, pdfborder={0 0 0}]{hyperref}
\usepackage{float}
\usepackage{dcolumn}
\usepackage[nolist,nohyperlinks]{acronym}
\usepackage{xspace}
\usepackage[abs]{overpic}
\usepackage{pict2e}
\usepackage{enumitem}
\usepackage[usenames,dvipsnames]{xcolor}
\usepackage[utf8]{inputenc}
\usepackage{acronym}
\usepackage{gensymb}
\usepackage{cleveref}
\usepackage[normalem]{ulem}
\usepackage{longtable}
\usepackage{verbatim}
\usepackage{multirow}
\usepackage{amsmath,amssymb}

\usepackage{subfigure}
\usepackage{placeins}
\usepackage{hyperref}
\usepackage{soul}

\newcommand{\vX}{\vec{X}}

\definecolor{rb4}{HTML}{27408B}

\begin{document}

	\title{Adaptive Kernel Density Estimation proposal  in gravitational wave data analysis}
	
   \author{Mikel Falxa }
   \affiliation{Université Paris Cit\'e, CNRS, Astroparticule et Cosmologie,  F-75013 Paris, France}
	
	\author{Stanislav Babak}
	   \affiliation{Université Paris Cit\'e, CNRS, Astroparticule et Cosmologie,  F-75013 Paris, France}
	 
	 \author{Maude Le Jeune}
	 \affiliation{Université Paris Cit\'e, CNRS, Astroparticule et Cosmologie,  F-75013 Paris, France}

\date{September 2021}

\newcommand{\comm}[1]{{(\bf Comment: #1)}}

\begin{abstract}
    Markov Chain Monte Carlo approach is frequently used within Bayesian framework to sample the target posterior distribution. Its efficiency strongly depends on the proposal used to build the chain. The best jump proposal is the one that closely resembles the unknown target distribution, therefore we suggest an adaptive proposal based on Kernel Density Estimation (KDE). We group parameters of the model according to their correlation and build KDE based on the already accepted points for each group. We adapt the KDE-based proposal until it stabilizes. We argue that such a proposal could be helpful in applications where the data volume is increasing and in the hyper-model sampling. We tested it on several astrophysical datasets (IPTA and LISA) and have shown that in  some cases  
    KDE-based proposal  also helps to reduce the autocorrelation length of the chains. The efficiency of this proposal is reduces in case of the strong correlations between a large group of parameters. 
\end{abstract}

\maketitle

\section{Introduction}

We live in the era of large physics and astrophysics projects and often have to deal with large and complex datasets. The data analysis usually requires large computing facilities and a single computation could sometimes last for weeks. 
Optimizing the analysis techniques and pipelines is then a key challenge of the data science associated with all large (astro)physical experiments. 

Nowadays, it is quite common to use Bayesian framework for analysing the data. It is especially convenient if we have a parameterized data model (or several competing models)  describing the data. In this approach we treat all parameters as random variables with some prior based either on some physical principles or informed from the previous independent experiments.  We use the observations at hands to refine our prior knowledge and infer a posterior probability distribution function for parameters of a model, or even perform a  selection among several  models.  Often we have to deal with multidimensional parameter space with a non-trivial likelihood function which can be evaluated only numerically. One of the most used tool to perform the numerical sampling from a target distribution is Markov Chain Monte Carlo (MCMC). Building a Markov chain that represents the desired posterior requires two key ingredients:  (i) proposal suggesting how to choose point $\vX_{i+1}$ given the last point in the chain $\vX_i$; (ii) the detailed balance which ensures the reversibility of the chain.  One of the most successful and frequently use proposal approach is to use parallel tempering: running several chains with logarithmic distributed temperature ladder (see for example \cite{Earl_2005}).  The hot chains play role of a proposal there and its efficiency depends on the interplay of number (and distribution) of hot chains (more chains is better) and computational demands (increase with the number of chains).  Understanding the properties of the signal and the likelihood surface often leads to a  custom proposal  suitable for a particular problem.

Here we suggest another generic proposal based on the  Kernel Density Estimation (KDE). The idea of using KDE is not new and might seem a rather trivial. The main result of this paper is in a particular implementation of KDE itself and its embedding into a sampler. Even though the proposed method is very generic we will mainly discuss its implementations in the gravitational waves (GWs) data analysis. 

Let us summarize the key points of the KDE-based proposal. 

\begin{itemize}

	\item  KDE is used together with the set of other  proposals in building a Markov chain. We assume an adaptive approach where we use the 
	data accumulated in a chain to regularly rebuild the KDE. We repeat adjustments until the convergence criteria based on the Kulback-Leibler divergence is satisfied. 

	\item In order to build KDE we split all parameters into several groups, where parameters in each group show evidence of mutual correlation.  The KDE-based proposal  is most efficient if  the full parameter space could be 
	split into many small uncorrelated groups. The performance drops significantly if dimensionality of a group is larger than 5. 
	\item We have build KDE  with the self-optimizing  bandwidth based on the distribution of a sample points provided at the input.
\end{itemize} 

Note that the adaptation breaks ``Markovian'' properties of the chain. Either one should dismiss the parts of the chain 
during the adaptation or, in case of uninterrupted adaptation,  assume  that the chain is only asymptotically Markov. 
We give detailed description of implementation in the next two sections (\ref{sec:kde}, \ref{sec:method}).

%a based method to help Monte Carlon Markov Chain (MCMC) algorithms. The idea is to have a KDE proposal distribution that will adapt itself to the currently produced chains as the MCMC is sampling. Having an adaptive proposal distribution allows us to explore the parameter space of a given model more efficiently without prior knowledge of its posterior distribution stucture. If the proposal distribution works well, we increase the number of independent samples and effectively reduce computation time.

We have implemented the KDE-based proposal in a particular sampler \url{https://gitlab.in2p3.fr/lisa-apc/mc3}. We give  a detailed description  of this sampler in Appendix \ref{sec:appendix_sampler}. The main feature of this sampler is that it runs several chains either completely independently or as parallel tempering. Multi-chain run is used to compute Gelman-Rubin ratio  \cite{10.1214/ss/1177011136} to monitor the convergence.  

We assess the performance of the suggested proposal in two applications to GW  data analysis. In  first one we analyse the data combined by International Pulsar Timing Array (IPTA) collaboration searching for a continous GW signal in the nano-Hz band. As the second dataset we use simulated LISA data publically available through LISA Data Challenge (2a) portal. We use the KDE-based proposal to infer parameters of 6 Galactic white dwarf binaries. 
  We present the performance of our proposal for those two data analysis problems in Section \ref{sec:results}, in particular we show that the KDE-based proposal allows to reduce the autocorrelation length while keeping high acceptance rate.

We conclude the paper with  discussion on the limitation and possible extension of our method in Section \ref{sec:conclusion}.   

%We want to make an easy to use MCMC sampler python library using \texttt{Numpy} \cite{numpy} and \texttt{Scipy} \cite{scipy} with several built-in proposal distribution amongst which the adaptive KDE. We will test it on astrophysical datasets of IPTA DR2 \cite{ipta_antoniadis_2022} and simulated LISA data (SANGRIA ?) to have examples of performance on high dimensional and complex parameter space.

\section{Kernel Density Estimation}
\label{sec:kde}

In this rather short section we describe our particular way of building  KDE. We start with a short introduction to KDE and then give details of the bandwidth  optimization that we use.

\subsection{Brief introduction}

KDE is a non-parametric method used to estimate a probability density function (pdf) based on a finite set of sample points  \cite{kde_0}\cite{kde_1}. It is a smooth alternative to a histogram. The advantage of KDE is that it uses no binning and gives a continuous function interpolating (and extrapolating) across the whole parameter space. For a d-dimensional dataset $\{\vec{X}\}$ of size $N$ and kernel $K(\textbf{x}, \vec{h})$, we have our KDE $\hat{f}(\textbf{x}, \vec{h})$ :

\begin{equation}
    \hat{f}(\textbf{x}, \vec{h}) = \frac{1}{N}\sum _{a=0} ^{N-1} K(\textbf{x}-\vec{X}_a, \vec{h}),
\end{equation}
with parameter $\vec{h}$ specifying the bandwidth of the kernel. We use latin subscripts from the first half of the alphabet to enumerate the samples in the set. 
The main idea is to sum smooth kernel functions of $\textbf{x}$ centered on each sample (input) data point $\vec{X}_a$. The overlaps between neighbouring kernels will add-up, shaping the PDF for the set of samples $\{\vec{X}\}$. The choice of the kernel is arbitrary and we choose to work with a gaussian kernel of the form:

\begin{equation}
    K_{g} (\textbf{x}-\vec{X}_a, \vec{h}) = \prod_{i=1} ^d \frac{\exp{\left\{-\frac{1}{2} \frac{|\textbf{x}-\vec{X}_a|^2_i}{h^2_i}\right\}}}{\sqrt{2\pi} h_i},
\end{equation}
where the $h_i$ is the local bandwidth corresponding to $i$-th parameter $|\textbf{x}-\vec{X}_a|_i$, and $d$ is the dimensionality of the parameter space.  We use the latin letter from the second half of the alphabet to enumerate particular parameters, and the vector notation corresponds to a vector in the parameter space.

\subsection{Optimal bandwidth}
\label{sec:opt_bw}

A KDE has one free parameter which we want to tune, the bandwidth $\vec{h}$. Its value should be adapted to the dataset we are working with. There is no direct way of estimating it and we use an optimisation method \cite{optimal_bw} based on the  minimisation of the mean squared error (MSE) $\epsilon ^2$ with respect to $\vec{h}$ :

\begin{equation}
    \epsilon ^2 = \int d\textbf{x} (\hat{f}(\textbf{x}, \vec{h}) - f(\textbf{x}))^2,
\end{equation}

\begin{equation}
    \frac{\partial \epsilon^2}{\partial \vec{h}} = 0,
    \label{eq:mse}
\end{equation}
where $f(\textbf{x})$ is the true pdf that we want to approximate with the KDE.
The numerical way of evaluating this integral is given in the appendix~\ref{A:optimband}. 

%\subsection{Local bandwidth}

Instead of using a global bandwidth $\vec{h}$, we can define a local bandwidth $\vec{h}_a$ for each kernel $K(\textbf{x}-\vec{X}_a, \vec{h}_a)$ \cite{local_bw}. In that case, our KDE $\hat{f}(\textbf{x}, \vec{h})$ is :
\begin{equation}
    \hat{f}(\textbf{x}, \vec{h}) =  \frac{1}{N}\sum _{a=0} ^{N-1} K(\textbf{x}-\vec{X}_a, \vec{h}_a).
\end{equation}
Intuitively we expect the local bandwidth $\vec{h}_a$ to be scaled according to the local density of points. Indeed, the bandwidth is chosen so that it is narrow in the regions of parameter space where the samples are most dense and it is broad where we have fewer samples. This can ensure good interpolation and overlap between kernels, in particular, in high dimensional problem where the sample points are very sparse.

In practice we use solution of equation \ref{eq:mse} to find optimal local bandwidth. The global bandwidth (if needed)
could be defined as an average over all local values (by setting the input parameter $\texttt{global bw = True}$). 

In Appendix \ref{A:optimband}, we have  shown that each local bandwidth $\vec{h}_a$ = [$h_{a, 1}$  $h_{a, 2}$  ...] of the kernel centered on point $\vec{X}_a$ can be approximated by solving a linear system for the $k$ nearest neighbours $\vec{X}_b$ of the form :

\begin{equation}
	\textbf{A}(\vec{X}_b) \begin{bmatrix}
		\frac{1}{h^2_{a, 1}} \\
		\frac{1}{h^2_{a, 2}} \\
		\vdots
	\end{bmatrix}
	=
	\vec{B}(k)
\end{equation}

Matrix $\textbf{A}$ and vector $\vec{B}$ are  given in Appendix \ref{A:optimband} eqn. \ref{eq:opt_bw_lin_system}, where we provide  a detailed description of the method. They depend on the position of 
 nearest neighbours  which are the $k_{near}$ points contained in a hypercube centered on the point $\vec{X}_a$ in the parameter space. The edge $\Delta X_i$ of the hypercube for each parameter $X_i$ is defined as :

\begin{equation}
	\Delta X_i = ( \max_{\{\vec{X}\}}{X_i} -  \min_{\{\vec{X}\}}{X_i}) / s
\end{equation}

where maximisation and minimisation are performed over the set of input samples  and $s$ is a scaling parameter we call "\texttt{adapt scale}".

It might happen that a hypercube contains no point (besides the central). In that case, we cannot compute the local bandwidth and the point is discarded. Its bandwidth is later set to the global bandwidth as defined above.
In case of high dimensionality and if the points in the dataset are very sparse, we change parameter $s$ iteratively decreasing by a factor 2 until we find non-empty hypercubes. However, if this happens the evaluation of the bandwidth will probably be flawed and there is not much we can do about it, except use bigger datasets with more sample points. Often the amount of additional points that is needed to cover all ``holes''  could be very large incurring  unmanageable  computational cost. That is why good parameter grouping is essential : reducing dimensionality without loss of correlated features in the data. This will be the main subject of the next section.

%Sometimes, we even have none of the hypercubes that contain points. In that case we can't compute the global bandwidth. It happens when dimensionality is too high and points in the dataset are way too sparse. To tackle the problem, the parameter $s$ will iteratively decrease by a factor 2 until we find non-empty hypercubes. However, if this happens the evaluation of the bandwidth will probably be flawed and there is not much we can do about it, except use bigger datasets with more points. But the amount of additional points we should use is tremendous, as will be the computational cost. This is why good parameter grouping is essential : reducing dimensionality without loss of correlated features in the data.

\section{Method}
\label{sec:method}

The main idea is to build a KDE for a given d-dimensional set of sample points $\{\vec{X}\}$. However, for a high dimensional KDE we are strongly affected by the "curse of dimensionality" \cite{curse_of_dim} because the sample sets are often limited in size leaving under-covered regions of the parameter space. In addition the efficiency of KDE is degrading if there are too many points since we place the kernel on top of each sample.  For that reason, we will split a d-dimensional parameter space into several low dimensional subspaces  grouping the most correlated parameters together. We assume that sub-groups are not correlated and build the KDE for each of them 
$\hat{f}_{\alpha} (\textbf{x}_{\alpha}, \vec{h}_{\alpha})$, so the total KDE is the product of the low-dimensional KDEs:
\begin{equation}
    \hat{F}(\textbf{x}, \vec{h}) = \prod _{\alpha} \hat{f}_{\alpha} (\textbf{x}_{\alpha}, \vec{h}_{\alpha}),
\end{equation}
where the greek indices enumerate the subgroups of parameters.  Forming these  subgroups  relies on the assessment of the correlation between parameters based on the provided set of samples $\{\vec{X}\}$ and this is the main subject of the next subsection.

\subsection{Parameter grouping}

\begin{figure}[th!]
    \centering
    \includegraphics[width=0.5\textwidth]{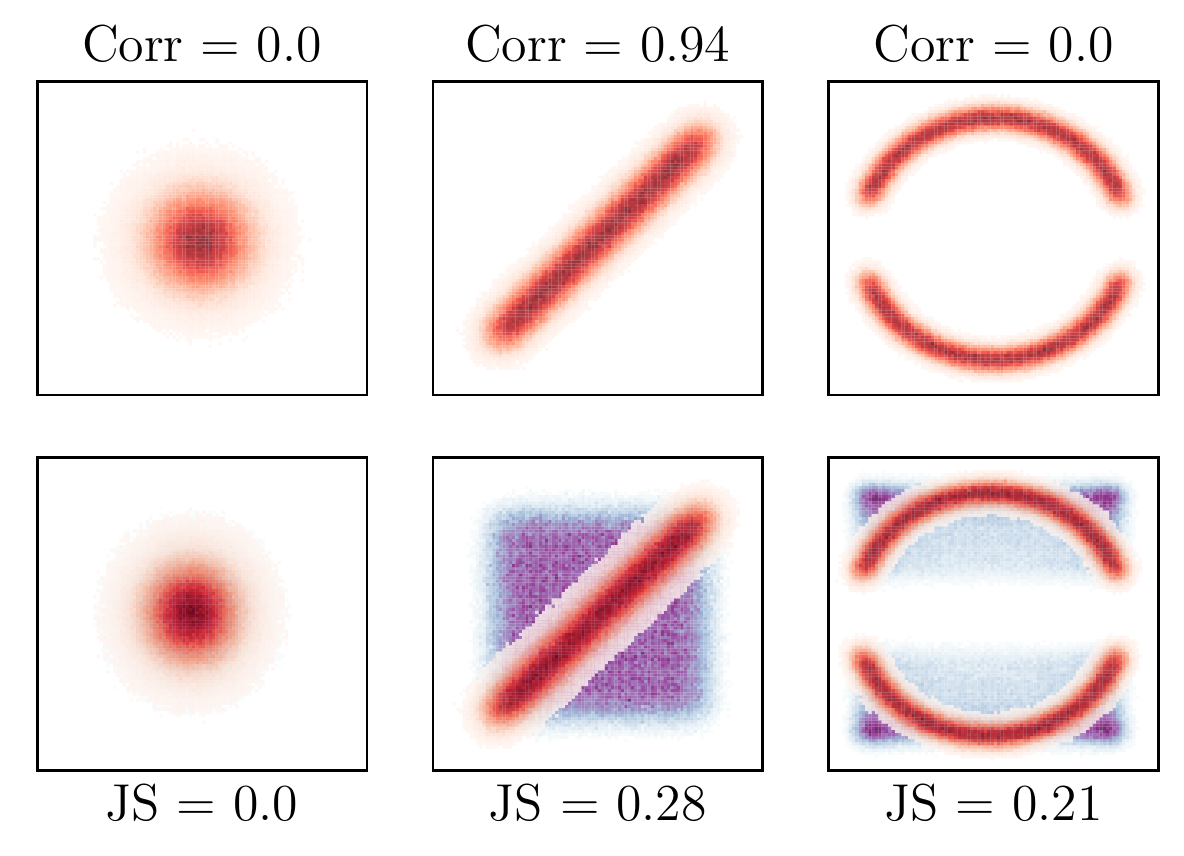}
    \caption{We plot three examples of datasets where we have on the left no correlation, in the middle linear correlation and on the right more elaborate features. On the top panels we have the corresponding values of the correlation coefficient based on simple evaluation of the covariance matrix. We see that it excels at finding the linear correlation but completely fails with the right panel features. On the bottom panels we have the same three datasets in red with their corresponding shuffled version that destroys correlations in blue. While the left panel remains unchanged, the others are affected and it is captured by the JS divergence.}
    \label{fig:js_vs_corr}
\end{figure}

For a d-dimensional dataset $\{\vec{X}\}$ we want to split the parameters in several sub-groups. Each sub-group will contain correlated parameters while parameters from  different sub-groups will be uncorrelated. We could use a 
covariance matrix to identify correlations, however it implicitly assumes Gaussian distribution and cannot account  for any complex 2D structures between pairs of parameters.  Instead we use a method based on the Jensen-Shannon divergence (JSD) \cite{js}. JSD, similarly to the Kullback–Leibler (KL) divergence, measures the similarity between two distributions, but with the advantage of being symmetric and bound $0 \le JSD \le \ln 2$, moreover it does not make any assumptions about the distributions.
 For each pair of parameter $(X_i, X_j)$  with the joined probability distribution $p(X_i, X_j)$ we compute 
 
 \begin{equation}
 	0 \leq \textrm{JSD}\big( p(X_i, X_j)||p(X_i)p(X_j) \big) \leq \ln 2,
 \end{equation}
 
 where the second distribution $p(X_i)p(X_j)$ is a product of one-dimensional (marginalized) distributions obtained by shuffling the parameters $\{X_{i}, X_{j}\}$  that supposed to destroy any correlations between parameters $X_i, X_j$ (see figure \ref{fig:js_vs_corr}).  If the JSD is low, it means that the shuffling did not affect the dataset and the parameters did not exhibit correlation. On the other hand, if the JSD is large, the shuffling did change something and the parameters should be grouped together.

\begin{figure*}[th!]
    \centering
    \includegraphics[width=1.0\textwidth]{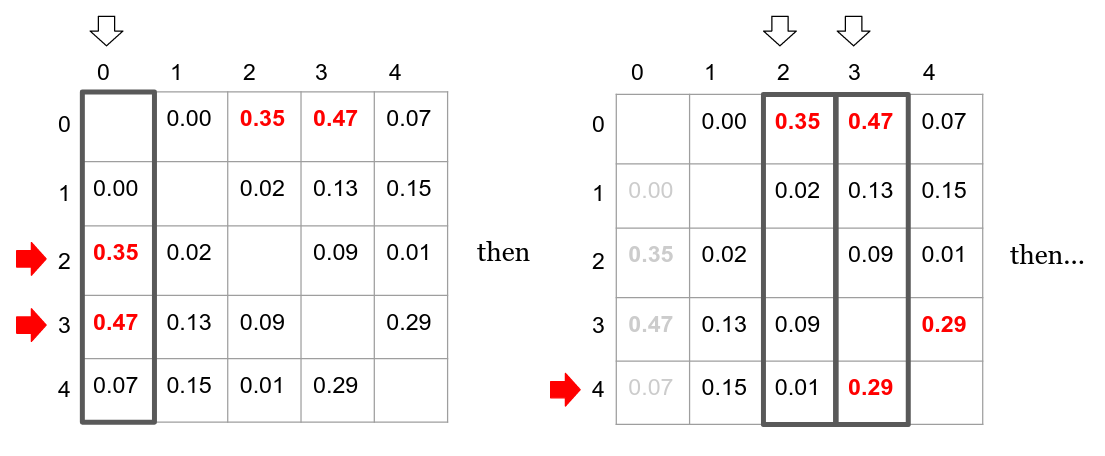}
    \caption{Illustration of the parameter grouping process using a JSD matrix. For this example, the JSD threshold is set to 0.25, hence the parameter sub-groups will be [0, 2, 3, 4] and [1]. Starting from parameter $0$, we find that JSD for parameters $2$ and $3$ are above the threshold so they are both correlated with parameter $0$. Then we check for parameters $2$ and $3$ and find that $4$ is correlated with $3$ while $2$ sees no additional correlation. The last step would have been to check for $4$ and find that there are no additional correlations. Therefore, $0$, $2$, $3$ and $4$ will  form a sub-group of correlated parameters. The parameter ``$1$'' is the last parameter that does not  correlate with others (according to the adopted threshold) and it will be a sub-group on its own.}
    \label{fig:algo_correlation_kde}
\end{figure*}

We define a JSD threshold (we usually use 0.1) which we consider that two parameters are correlated. Starting from one parameter, we iterate the process to extract all  correlated (chained) pairs.  Once no additional correlated parameter is found, we take the union of all correlated pairs of parameters to form a sub-group. This process is illustrated in figure \ref{fig:algo_correlation_kde}.

In case of multimodality of the probability distribution function that we try to reproduce with KDE, we implemented 
an additional (optional) feature: clustering samples before building KDE. This feature is especially welcome when the modes are separated by very low probability valleys. We cluster the samples (using k-means method  \cite{kmeans})
and apply KDE building approach described above to each mode. This does not change the fundamental structure of the KDE but it helps the bandwidth adaptation.

%\begin{figure}
%	\centering
%	\includegraphics[width=0.5\textwidth]{figures/clutser_paper_2.png}
%	\caption{An illustration of the clustering method. We first find 1D clusters, here there are 4 (0, 1, 2 and 3), then we mask the data to "roughly" isolate n-dimensional modes. The bandwidth is computed independently for each mask. Note that masks 03 and 12 contain very few points (less than 10\% of the total number of points). Therefore, their bandwidth will be set to the mean of other mask's bandwidths (here 02 and 12).}
%	\label{fig:clustering}
%\end{figure}
%
%
%We first count the number of modes in each binned 1D histogram $p(\vec{n})$ of parameter $\vec{n}$ with the $scipy$ function $scipy.signals.findpeaks$ \footnote{We have chosen to work with 1D histograms because clustering methods usually start to perform poorly at higher dimensions and become computationally expensive.}. Each peak position is given as initial centroid position for the $scipy.clutser.vq.kmeans2$ function. This way we isolate data points for each peak and each parameter. Because we are working with 1D posteriors, we use masks to isolate unions of 1D peaks that roughly correspond to $N$ dimensional modes in the dataset (see figure \ref{fig:clustering}). Sometimes, these masked subsets contain few or no points (less than 10\% of the total number of points in the dataset). In that case, their bandwidth will be set later to the mean bandwidth of non-empty modes obtained from non-empty masked subsets.

%\subsection{Adapting the bandwidth}

\subsection{KDE: turning posterior into proposal}

It is desirable in several application to use posterior points inferred for some parameters into a prior for another investigation.  Let us give several examples:

\begin{itemize}
	\item The data inferred from electromagnetic observations in form of  samples  is used as a prior for the 
	GW experiment. In this case we can either build a joined likelihood or, alternatively, build 
	KDE on the external posteriors and use it as a prior while analyzing the GW data.   
	\item In Pulsar Timing Array (PTA) data analysis we often first investigate data acquired for each pulsar and trying to build a noise model. Later this pulsar and associated noise model is plugged into "Array" of pulsars for searching for a GW signal. It is proven to boost significantly  efficiency of the GW search if we use posteriors for the noise  model inferred in the first step as a proposal in the global fit later on. 
	\item Often the data is taken continuously and we want to analyse it "on the fly"; that is true for GW data analysis. In this case you want to increment the data with a certain cadence while using information about the sources acquired from the analysis of  the past data. One possibility is to turn again posterior built from the analysis 
	of, say, first half a year of data into a prior in the analysis of the whole year of data.
	\item The method in \cite{hypermodel} gives a practical suggestion on how to compute the Bayes factor comparing several models without computing the evidence for each model. In this approach we introduce hyper-parameter indexing the models and jump in this parameter (say, within MCMC) which corresponds to jumping between the models. For this method to be robust the exploration within each model must be very efficient otherwise it will lead to very long poorly converging runs or to spurious results.  If we have posteriors for each (or some) model available, we can turn them into proposal and use in the hyper-model  exploration.
\end{itemize}

%\subsection{Adapative KDE}
%
%The purpose of this KDE is to be used as a proposal distribution with MCMC samplers \cite{hastings}. It can be used as a fixed proposal, or alternatively as an adaptive proposal. In the latter case, we want our KDE to continuously adapt to the currently sampled dataset. It would allow us to have an appropriate proposal distribution without prior knowledge of our posterior distributions.
%
%\subsubsection{KDE proposal}

The main idea of grouping parameters in building KDE implies that we probably also want to make jumps within each subgroup (or in some subgroups) while keeping other parameters fixed (Gibbs-like sampling). The subgroups for a current jump are chosen randomly assuming equal probability attached to each subgroup.  The randomness implies reversibility and equal probability is a warrant that we jump in all parameters evenly (on average)  while performing low dimensional jumps. Let us denote the number of sub-KDEs that is used for each jump $n_{kde}$, then the proposal probability is 

\begin{equation}
	\hat{F}_{n_{kde}}(\textbf{x}, \vec{h}) = \prod _{\alpha} ^{n_{kde}} \hat{f}_{\alpha} (\textbf{x}_{\alpha}, \vec{h}_{\alpha}),
\end{equation}
where the subscript in $\textbf{x}_{\alpha}$ implies that  we vary only parameters that belong to that ($\alpha$) subgroup. This probability is used to balance the chain in the Metropolis-Hastings step of the MCMC algorithm \cite{hastings}.  Choosing a point from a given sub-KDE $\hat{f}_{\alpha} (\textbf{x}_{\alpha}, \vec{h}_{\alpha})$ is done by drawing a point from the randomly chosen kernel $K(\textbf{x}-\vec{X}_a, h_a)$ of $\hat{f}_{\alpha} (\textbf{x}_{\alpha}, \vec{h}_{\alpha})$ centered on $\vec{X}_a$:
\begin{equation}
	\vec{X}_b^*  \rightarrow \vec{X}_a + \mathcal{N}(\vec{0}, h_a),
\end{equation}
where $\mathcal{N}(\vec{0}, h_a)$ is a normally distributed random variable with $\vec{0}$ mean and covariance matrix $\textrm{diag}(h_a)$ that is the bandwidth of kernel $K(\textbf{x}-\vec{X}_a, h_a)$. For a random set of  $n_{kde}$ sub-KDEs, the newly proposed point $\vec{X}^*$ is the union of  parameters from each subgroup $\vec{X}_b^*$ :

\begin{equation}
	\vec{X}^* = \bigcup _b\vec{X}_b ^*.
\end{equation}

\subsection{Adaptive proposal}

In case we do not have samples from the previous investigations, we still can build KDE-based proposal using the points accepted by a running MCMC. There are several caveats which need to be considered: (i) during the burn-in and even some time after the distribution of the accepted points is quite unstable that will reflect on the KDE (ii) we are breaking the rules of MCMC, the chain is only asymptotically Markov,  so that at some point we should fix the KDE-based proposal and dismiss all samples accumulated before.  The rest of this subsection gives the details of the practical implementation of the adaptation.

\begin{figure}
    \centering
    \includegraphics[width=0.5\textwidth]{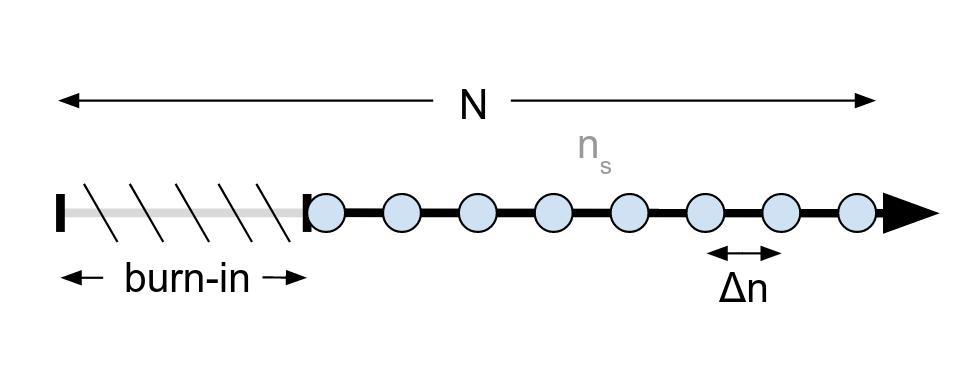}
    \caption{For a chain of total length $N$, we get rid of the burn-in, then we extract $n_s$ linearly space samples from the remaining fraction of the chain. These $n_s$ points are used to build the KDE.}
    \label{fig:adapt_chain}
\end{figure}

To build a KDE on the currently sampled chain, we select only a subset of the total $N$  samples. In particular, we take $n_s$ uniformly spaced points, from burn-in to the last point, as illustrated in Figure \ref{fig:adapt_chain}. The burn-in is taken to be a fraction, $q$, of the total chain length $N$  and those points are dismisssed (hence, $n_s = (1-q) N$).
As chain evolves, the burn-in could also grow until we get to the stationary distribution. The post-burn-in length is also increasing, and, since we keep $n_s$ fixed,  the space $\Delta n$ between the selected samples grows. We expect that the quality of the KDE improves with increase in $\Delta n$ because of the reduced correlation in the samples taken for building KDE, and that is true until $\Delta n$ reaches typical autocorrelation length of the chain \cite{mcmc_autocorr}.

We re-build a new KDE after each $N_{adapt}$ iterations (jumps) of the chain. We want to track the evolution of KDE and stop adapting when it has reached stability (that could be another indicator of the burn-in phase).  We compare the new (re-built) KDE $\hat{F}_1(\textbf{x}, \vec{h_1})$ with the old $\hat{F}_0 (\textbf{x}, \vec{h_0})$
by computing the KL divergence \cite{kl}:

\begin{equation}
    \textrm{KL}(\hat{F}_0||\hat{F}_1) = \int d\textbf{x} \hat{F}_0(\textbf{x}) (\log \hat{F}_0(\textbf{x}) - \log \hat{F}_1(\textbf{x}))
    \label{eq:kl_def}
\end{equation}
This integral could be approximated as

\begin{equation}
    \textrm{KL}(\hat{F}_0||\hat{F}_1) \simeq \frac{1}{n_s} \sum _a (\log \hat{F}_0 (\vec{X}_{0a}, \vec{h_0}) - \log \hat{F}_1 (\vec{X}_{0a}, \vec{h_1})),
    \label{eq:kl}
\end{equation}
where  $\{\vec{X}_0\}$ is the set of  $n_s$  samples used to build $\hat{F}_0 (\textbf{x}, \vec{h_0})$.  This gives a measure of  change in KDE between successive  updates: $\Delta \textrm{KL}$. 
 We stop updating  KDE  in order to preserve the ergodicity of the process \cite{adaptive_mcmc}  as soon as convergence criterion
\begin{equation}
    \frac{|<\Delta \textrm{KL}>|}{\sqrt{<\textrm{KL}^2>}} < 5\%
    \label{eq:dkl_convergence}
\end{equation}
is satisfied. The angular brackets denote the averaging over the last 5 updates and we demand that the average change in KL is small compared to the average KL values. $\Delta$KL can be negative or positive, depending on the evolution of KL. If KL does not converge to a specific value, this condition ensures that it is at least oscillating around a mean value.

\section{Results}
\label{sec:results}

% Test on PTA and LISA data :\\
% KDE QUALITY :\\
% - Show the groups that were found, illustrating them with corner plots. (76x1, 12x2, 1x4)\\
% - Show overplot corner plot of original data against redrawn dataset from KDE.\\
% - Try to compute KL using binning, with original dataset against redrawn dataset from KDE. Check KL % for whole dataset against whole KDE.\\
% - Also check individual sub-KDEs KLs, plot them in a histogram. Could justify the sub-KDE by sub-KDE % jumps.\\
% KDE PROPOSAL QUALITY :\\
% For same model, 3 different runs with KDE proposal, binned proposal, no proposal :\\
% - compare acceptance rate\\
% - compare autocorrelation length\\
% ADAPTIVE KDE PROPOSAL QUALITY :\\
% Rerun previous model with adaptive KDE:\\
% - compare acceptance rate\\
% - compare autocorrelation length\\
% - check KL and dKL evolution to show stopping criterion validity

We consider two datasets and perform search/parameter estimation using MCMC with KDE-based proposal. 

In first application we consider a dataset from International PTA collaboration and perform the noise analysis for each pulsar in the array \cite{Perera2019sca}. The likelihood is expected to be quite broad and unimodal but the dimensionality of the parameter space is large as well as its overall volume. 

In second application we work with the simulated LISA data and search/characterize small bandwidth with several Galactic white-dwarf binaries. The likelihood in this case has more complex structure with quite strong correlation between parameters. 

 We quantify performance of the KDE-based proposal using the following criteria:

\begin{itemize}
    \item  Closeness between built KDE and the true distribution using \textbf{KL divergence}\footnote{Because we do not know the true distribution, we have to use histograms to evaluate KL. The formula for binned KL is given in Appendix \ref{A:kl}.}.
    \item The \textbf{acceptance rate} when the KDE is used as a proposal with MCMC sampler
    \item The \textbf{autocorrelation length} \cite{mcmc_autocorr} of the MCMC chain when the KDE is used as proposal, to evaluate improvement in the mixing of the chain.
\end{itemize}

\subsection{IPTA dataset}

\begin{figure}
    \centering
    \includegraphics[width=0.5\textwidth]{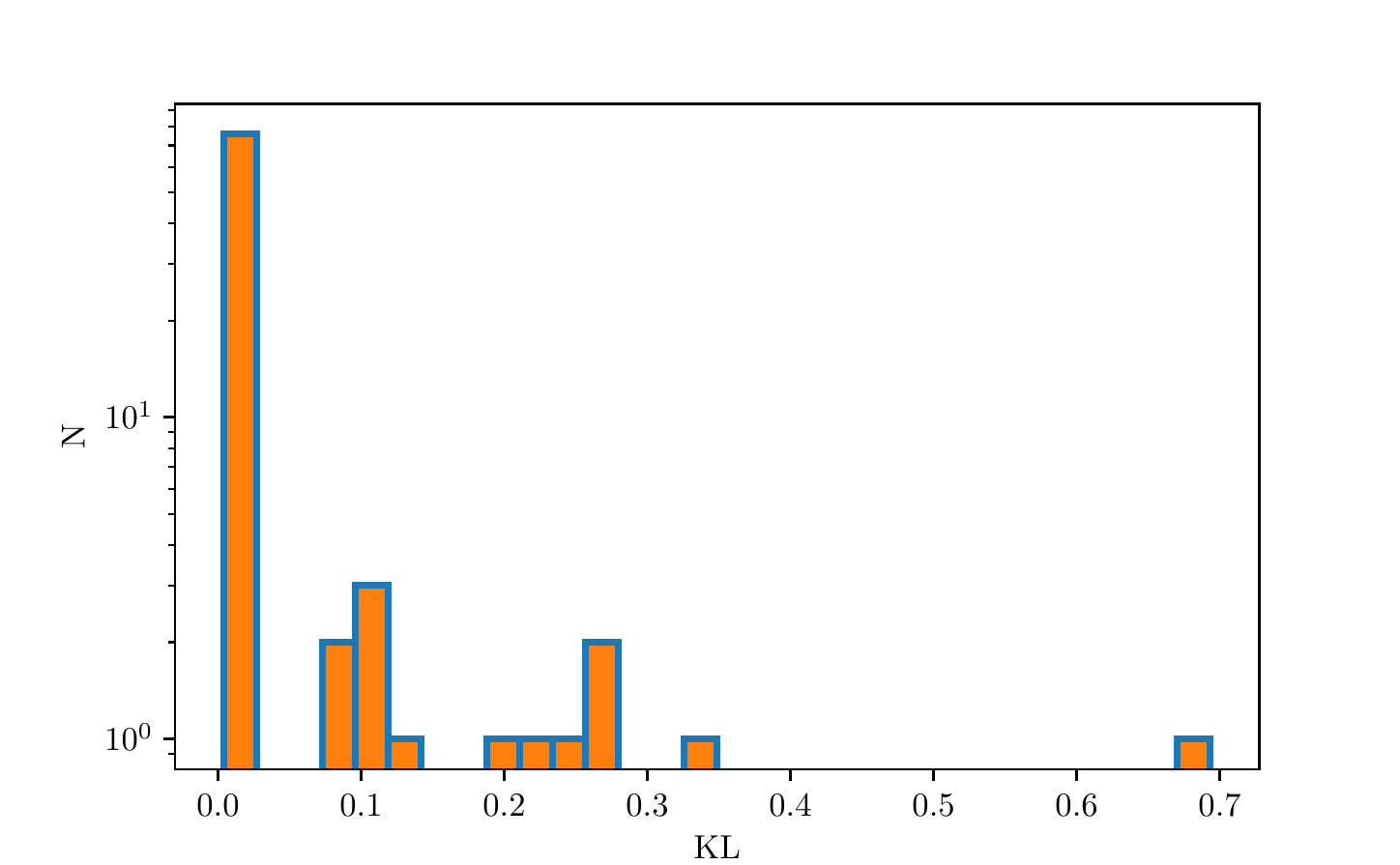}
    \caption{Distribution of the KL divergence for all sub-KDEs. 1-d sub-KDEs are of best quality.}
    \label{fig:KL_hist}
\end{figure}

We  build the KDE using $n_s = 10000$ samples from  the chains generated by previous MCMC runs. We adopt the following choice of parameters for generating KDE:

\begin{itemize}
    \item \texttt{js threshold = 0.1}
    \item \texttt{adapt scale = 10}
    \item \texttt{use kmeans = False}
    \item \texttt{global bw = True}
    \item \texttt{n kde = 1}
\end{itemize}
Within total 104 parameters, the grouping algorithm finds : \textbf{76} 1-dimensional sub-groups, \textbf{12} 2-dimensional sub-groups and \textbf{1} 4-dimensional sub-group. For each of these subgroups we calculate the KL divergence $\textrm{KL}(p_\alpha (\textbf{x}_\alpha)||\hat{f}_\alpha (\textbf{x}_\alpha, \vec{h}_\alpha))$ of the true pdf of the subgroup $p_\alpha (\textbf{x}_\alpha)$ against the corresponding sub-KDE $\hat{f}_\alpha (\textbf{x}_\alpha, \vec{h}_\alpha)$. A good sub-KDE should give a $\textrm{KL}(p_\alpha (\textbf{x}_\alpha)||\hat{f}_\alpha (\textbf{x}_\alpha, \vec{h}_\alpha))$ that is close to 0. Because the number of parameter is large, we show the histogram plot in figure \ref{fig:KL_hist} depicting KL for each subgroup.

\begin{figure}
    \centering
    \includegraphics[width=0.5\textwidth]{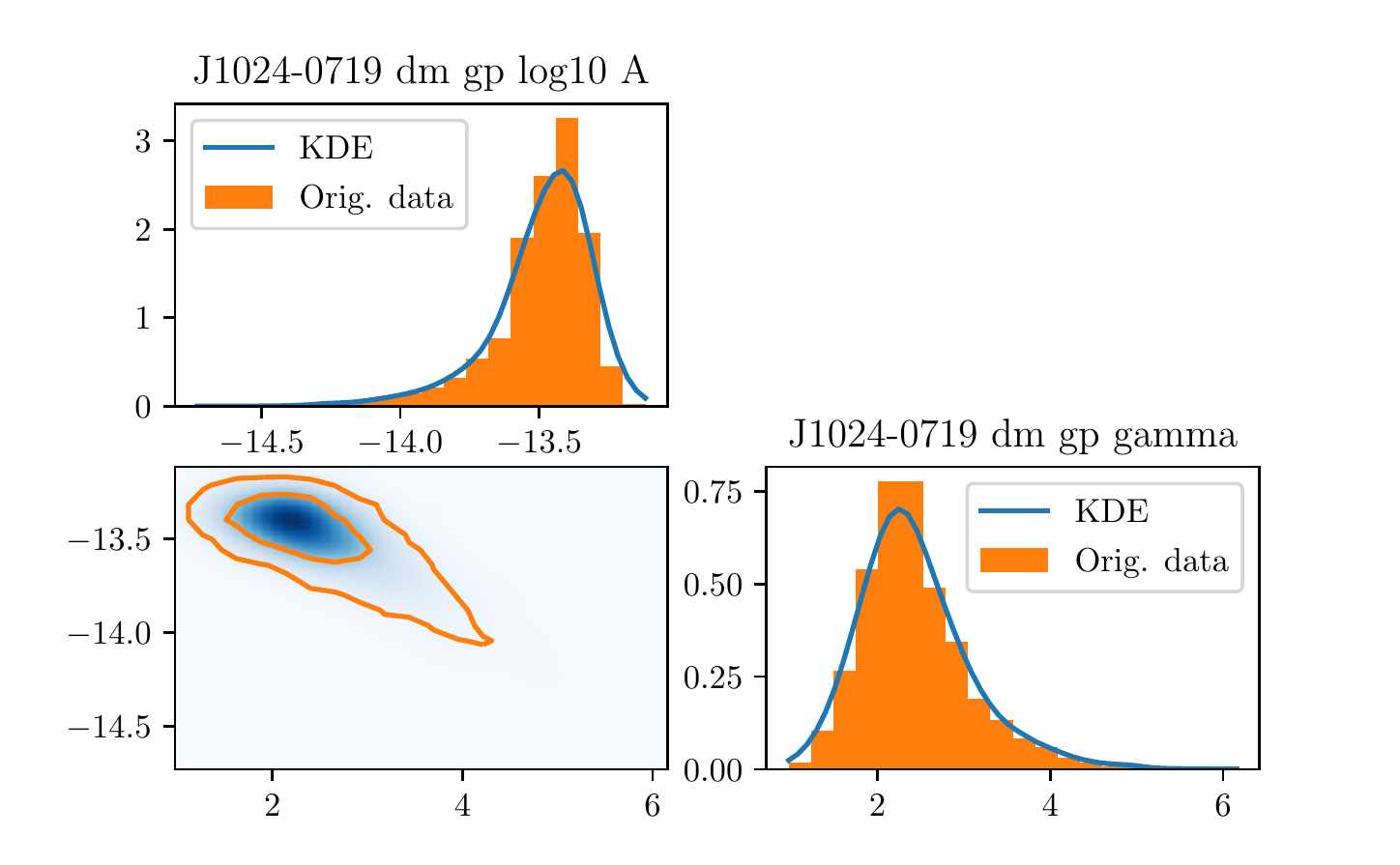}
    \caption{2-dimensional corner plot of binned original dataset against smooth KDE plot. KDE is blue, original is orange. The contour levels for the original dataset on the bottom left panel are [0.1, 0.95]. For this sub-KDE, $\textrm{KL} \simeq 0.11$.}
    \label{fig:2d_corner}
\end{figure}

The impact of the dimensionality of the sub-KDE on KL can be clearly seen in Figure~\ref{fig:KL_hist} . Close to 0 we have  76 1-dimensional sub-KDEs, between 0.1 and 0.4 we have 12 2-dimensional sub-KDEs, and the 4-dimensional sub-KDE has $\rm{KL} \approx 0.7$.
 As discussed in the previous section,  increasing dimensionality implies sparse data samples, so we expect KL values to rise because the KDE might fail to interpolate the PDF correctly between neighbouring points, producing holes in the distribution.  Computing KL allows us to quantify this effect and assess the quality of the KDE that may not be obvious by just eyeballing (see figure \ref{fig:2d_corner}).

Next we will analyse the IPTA data using MCMC sampler \cite{ptmcmc} and ENTERPRISE \cite{enterprise}  package for computing likelihood function. We chose to use two jump proposals Single‐Component Adaptive Metropolis (SCAM) \cite{SCAM} and Differential Evolution (DE) \cite{Braak, braak_vrugt} to compare to KDE. 
We search for a continuous gravitational wave signal while fitting for pulsar noise parameters \cite{nanograv_cw_11yr}.  The KDE for the noise parameters has been built using posterior samples obtained from 
the preceding single pulsar analysis. We will compare three different runs:
\begin{itemize}
    \item using a KDE-based + default jump proposals SCAM and DE (labeled as ``KDE'');
    \item using binned empirical distributions + default proposals SCAM and DE (labeled as ``Binned'');
    \item using only default proposals SCAM and DE (labeled as ``None'').
\end{itemize}
The binned empirical distributions are essentially 2-dimensional histograms based on the same posterior samples as used in building KDE \cite{enterprise_extensions}. Those carry a similar spirit to KDE, being pair-wise approximation 
to marginalized posterior, but at the same time are fundamentally different from the KDE in the sense that the KDE is a continuous function in space that interpolates between the sample points using smoothening kernel.
 In addition KDE-based proposal makes grouping based on the parameter correlation that could lead to more than two dimensional group (see figure \ref{fig:KL_hist}).

\begin{table}
    \centering
    \begin{tabular}{c|c|c|c}
         & \textbf{KDE} & \textbf{SCAM} & \textbf{DE}\\
        \hline
        \textbf{Acceptance rate} & 0.57 & 0.39 & 0.43
    \end{tabular}
    \caption{Acceptance for various proposals (KDE run)}
    \label{tab:kde_acceptance}
\end{table}

\begin{table}
    \centering
    \begin{tabular}{c|c|c|c}
         & \textbf{Binned} & \textbf{SCAM} & \textbf{DE}\\
        \hline
        \textbf{Acceptance rate} & 0.47 & 0.41 & 0.43
    \end{tabular}
    \caption{Acceptance for various proposals (binned proposal run)}
    \label{tab:binned_acceptance}
\end{table}

\begin{figure}
    \centering
    \includegraphics[width=0.5\textwidth]{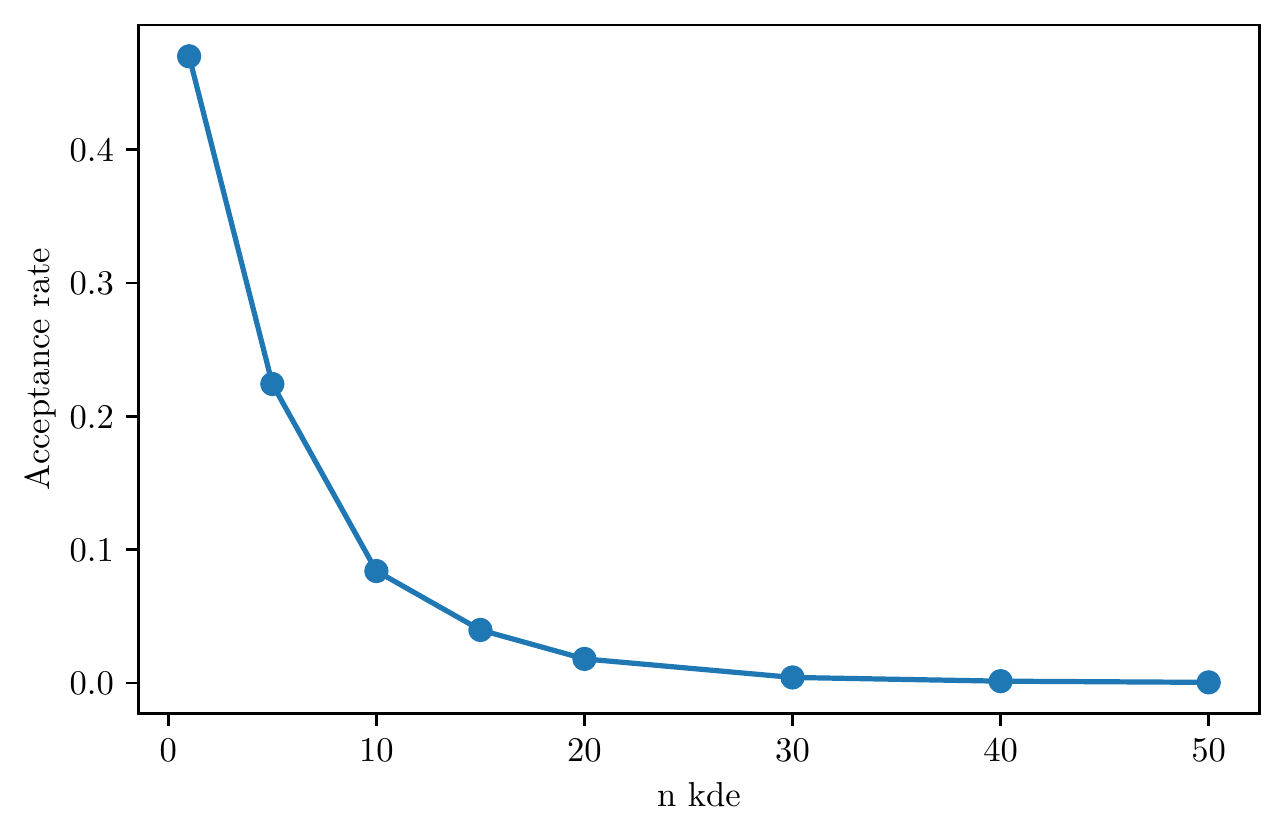}
    \caption{Acceptance rate for several values of $n_{kde}$.}
    \label{fig:accept}
\end{figure}

\begin{table}
    \centering
    \begin{tabular}{c|c|c|c}
        \textbf{$n_{kde}$} & \textbf{KDE} & \textbf{SCAM} & \textbf{DE}\\
        \hline
        \textbf{1} & 0.47 & 0.41 & 0.43 \\
        \textbf{5} & 0.22 & 0.36 & 0.41 \\
        \textbf{10} & 0.08 & 0.35 & 0.40
    \end{tabular}
    \caption{Acceptance for different $n_{kde}$}
    \label{tab:nmax_acceptance}
\end{table}
Tables~ \ref{tab:kde_acceptance},~\ref{tab:binned_acceptance},  compare acceptance rate of different proposals in two independent runs. One can see that  KDE has the highest acceptance rate due to smart parameter grouping and 
 interpolation between the samples   incorporated in the KDE-based proposal. 
For the jumps we could use simultaneous jump in one or several ($n_{kde}$) sub-groups.   Increasing $n_{kde}$ leads to the higher dimensionality of the jumps and  has a strong impact on the acceptance rate as shown in Figure~\ref{fig:accept}.  The best results are achieved if we perform jumps in one subgroup at the time, the acceptance rate decreases exponentially with $n_{kde}$. 

A high acceptance rate is not necessarily a sign of  a good proposal as it has to be paired with the low autocorrelation length. For each run, we compute the autocorrelation lengths of all parameters and compare the maximum, minimum and mean autocorrelation lengths. The low autocorrelation length implies that the samples drawn/accepted are independent. Results are presented in Table \ref{tab:autocorr}. KDE performs very well for IPTA data. 
Minimum autocorrelation length does not seem to be affected much by the choice of proposal but the maximum is reduced by a factor 2 when using KDE. Reducing the maximum is the most important because thinning the chain by this factor ensures that all samples are independent for all parameters. The mean autocorrelation length is just an indicator of the average performance of the proposal.

\begin{table}
    \centering
    \begin{tabular}{c|c|c|c}
         & \textbf{max} & \textbf{min} & \textbf{mean} \\
         \hline
        \textbf{KDE} & 578 & 27 & 113 \\
        \textbf{Binned} & 1386 & 29 & 160 \\
        \textbf{None} & 1032 & 25 & 208
    \end{tabular}
    \caption{The maximum, minimum and mean autocorrelation lengths for three runs.}
    \label{tab:autocorr}
\end{table}

\begin{table}
    \centering
    \begin{tabular}{c|c|c|c}
        $n_{kde}$ & \textbf{max} & \textbf{min} & \textbf{mean} \\
         \hline
        \textbf{1} & 578 & 27 & 113 \\
        \textbf{5} & 386 & 25 & 79 \\
        \textbf{10} & 395 & 28 & 95
    \end{tabular}
    \caption{The maximum, minimum and mean autocorrelation lengths for different $n_{kde}$.}
    \label{tab:nmax_autocorr}
\end{table}

\begin{figure}
    \centering
    \includegraphics[width=0.5\textwidth]{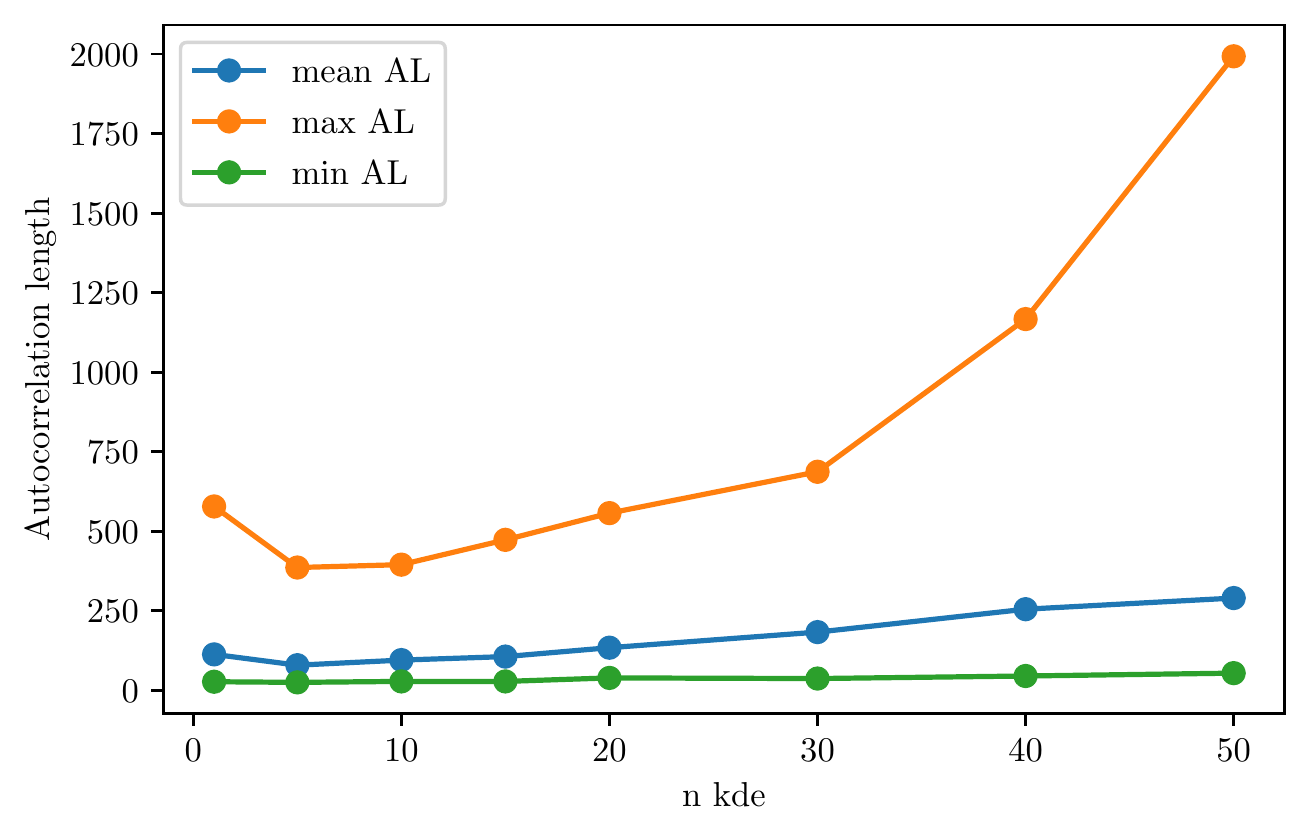}
    \caption{Mean, maximum and minimum autocorrelation lengths for several values of $n_{kde}$.}
    \label{fig:al}
\end{figure}

Like for the acceptance rate, we check the influence of $n_{kde}$ on the autocorrelation length. Results are given in Table \ref{tab:nmax_autocorr}. For high values of $n_{kde}$, even though we have decreasing acceptance rate,  the autocorrelation drops too and mixing improves. However, this result should be take with a caution, using high $n_{kde}$ could lead to a very low acceptance point as shown in Figure~\ref{fig:accept}. 
In Figure~\ref{fig:al}  we show the autocorrelation length as a function of  $n_{kde}$ and it indicates that the optimal number is around $n_{kde} = 5$ with acceptance rate of 0.22 that is very close to the expected optimal acceptance rate of 0.234 \cite{optimal_ar}. Based on these results,  we recommended to work with values of $n_{kde}$ between 1 and 10, especially when the dimensionality of the parameter space is high  like in this IPTA example.

\subsection{LISA dataset }

Now we turn our attention to the simulated LISA data, in particular, we use ``Sangria'' dataset which is 1 year long, it contains about a dozen of merging massive black hole binaries and about 30 millions of Galactic binaries 
\url{https://lisa-ldc.lal.in2p3.fr/}. Here we are interested in Galactic binaries in the very narrow frequency range around 4 mHz and we have removed all merging black holes. We have detected 6 sources in that frequency interval and we use KDE-based proposal together SCAM and DE\footnote{The DE introduced in \cite{Braak} is using population MCMC, here we rather use it on a single chain in the spirit described in the snooker proposal in \cite{braak_vrugt}}. This time we use home-made sampler $M3C2$ \url{https://gitlab.in2p3.fr/lisa-apc/mc3}, specifically parallel tempering version of it. We describe this sample in details in the Appendix ~\ref{sec:appendix_sampler}. This sampler is using Metropolis-Hastings acceptance-rejection step as well as \emph{slice} sampling \cite{slice}.

Each Galactic binary is characterized by 8 parameters, so we have in total 48 parameters \cite{Littenberg:2011zg}. We expect some parameters (like amplitude and the orbital inclination angle) to correlate for each source, and, in addition, some parameters could correlate between the sources. Here we try to build a proposal on-the-fly. The likelihood surface  for this problem is rather complex having many well separated maxima (reason for using parallel tempering). We will build KDE as we accumulate samples: adapting KDE proposal with the rate every 5000 samples and using  $n_s = 5000$ samples for each chain.  Besides KDE we also use SCAM proposal and slice sampling.  Note that the combination of SCAM jump with Metropolis-Hastings sampling and slice (being very independent) already significantly reduce the autocorrelation of the accepted points. 

%\comm{Need help here}
%We will test the adaptive KDE proposal on simulated LISA data (cite paper on simulated lisa). We will perform a search for 6 galactic binary sources (cite paper on model), making up for a 50 dimensional parameter space. This model produces complex posteriors that may be very challenging for our proposal. We set an update rate of $n_{update} = 5000$ (every 5000 samples) and we build KDEs with $n_s = 5000$ samples. We use the KDE proposal with the mc3 sampler (cite something), together with SCAM and SLICE proposal (cite something on slice). Like in the previous section, we check acceptance rate and autocorrelation length.

\begin{figure}[h!]
    \centering
    \includegraphics[width=0.5\textwidth]{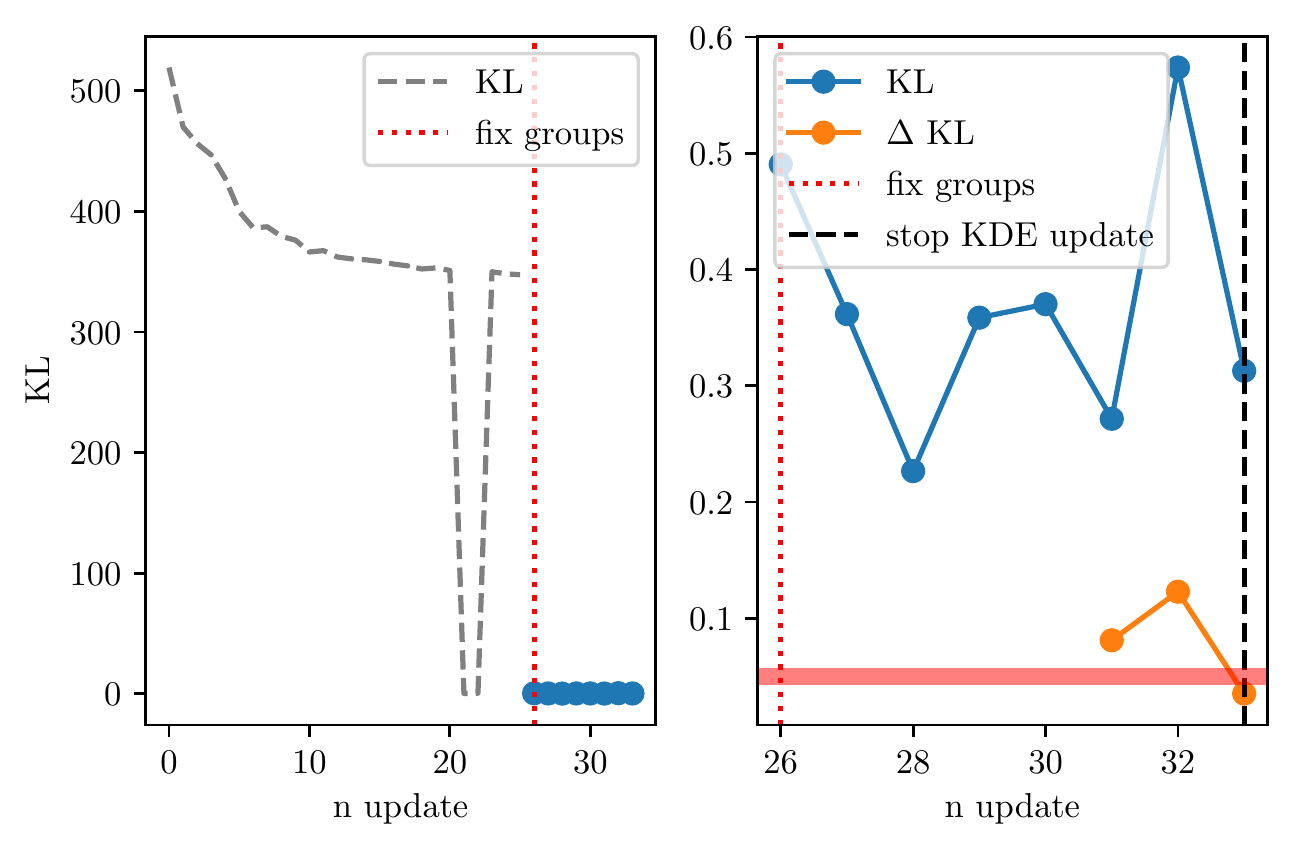}
    \caption{Evolution of KL and $\Delta \rm{KL}$. Left panel, evolution of KL before fixing parameter groups. Right panel, evolution of KL after fixing parameter groups,  we start computing $\Delta \rm{KL}$ 5 updates after the grouping was fixed. The thick red line indicates the $\Delta \rm{KL}$ threshold level of 5\% below which we reach convergence.}
    \label{fig:kl_nkde_1}
\end{figure}

First, we consider the convergence of KDE adaptation.  During the burn-in stage the KDE is changing quite violently. 
The correlation between parameters is quite unstable which leads to fluctuation in how parameters are grouped and in the number of subgroups. As burn-in proceeds we keep track of the grouping  and fix the splitting in sub-groups as soon as it stabilizes (when the same grouping appears at least 5 times).  Once we have fixed sub-groups we check the KL divergence between the subsequent updates of KDE (as described above).  The results of KDE adaptation are presented in Figure~\ref{fig:kl_nkde_1}. The consistent grouping of parameters was reached after 26 updates as indicated by a dashed red line.  Then we compute KL after each update and stop adaptation once the condition~\ref{eq:dkl_convergence} is met (see the right panel  of Figure~\ref{fig:kl_nkde_1}). From then on we keep KDE fixed and  perform the actual sampling.  

As a next step we want to check the acceptance rate of the KDE-based proposal and compare it to SCAM and slice. Note that slice is not based on the Metropolis-Hastings acceptance/rejection algorithm, however we still can introduce an effective acceptance rate as a ratio of total number of slice calls to the total number of likelihood evaluations used by slice. 

\begin{figure}[h!]
    \centering
    \includegraphics[width=0.5\textwidth]{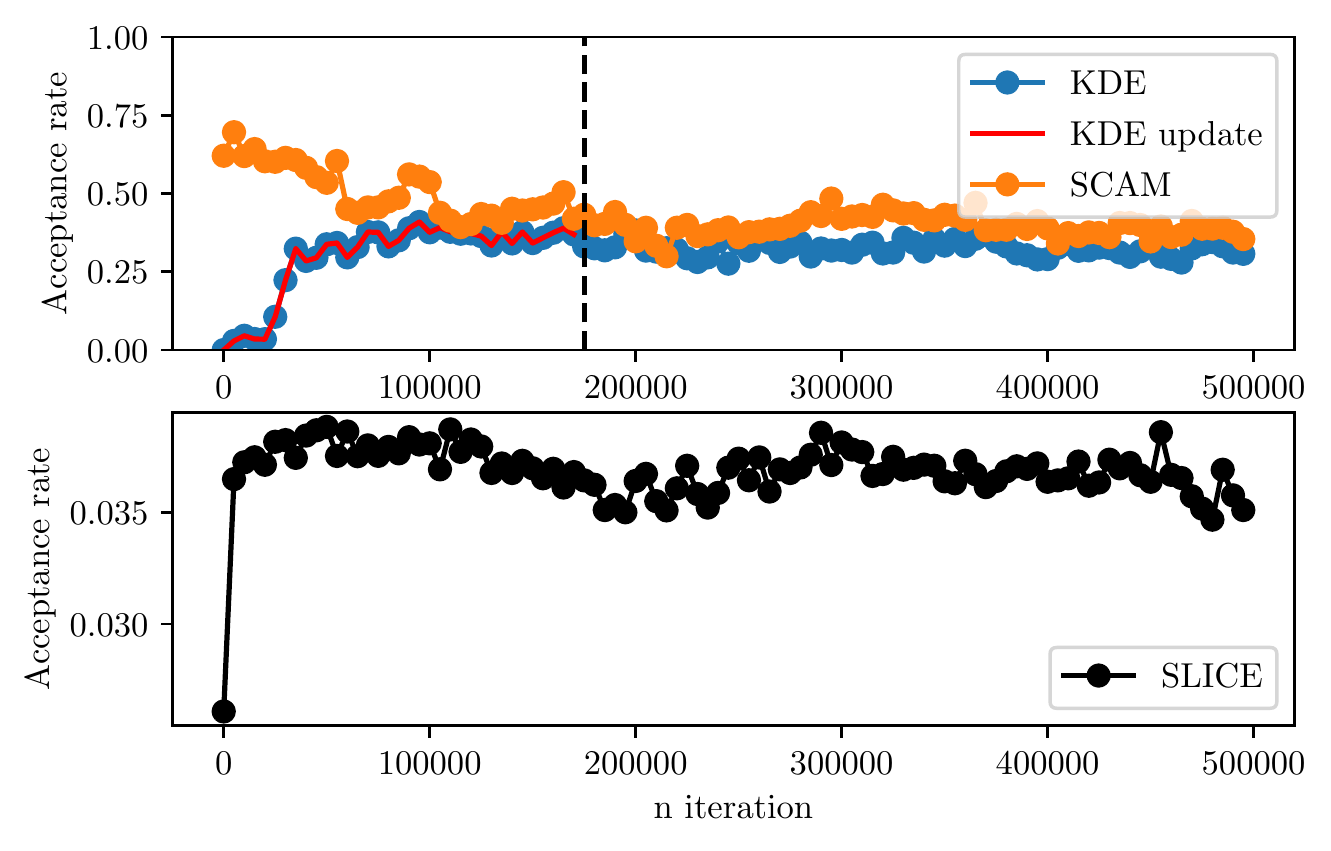}
    \caption{Acceptance rate of all proposals for $n_{kde} = 1$. KDE and SCAM are on top panel, SLICE on bottom panel. Red plot and black dashed line shows where KDE finally converged and stopped updating.}
    \label{fig:ar_nkde_1}
\end{figure}

We consider two cases for the subgroup jumps $n_{kde} = 1$ and $n_{kde} = 5$. 
Figure~\ref{fig:ar_nkde_1} compares acceptance rate for  $n_{kde} = 1$.  One can see that it stabilizes 
around 0.31 very fast, and, despite that it is lower than what we had for the PTA application, it is a very decent acceptance rate.  SCAM has a similar acceptance (but usually longer autocorrelation length), while slice is worse by a factor 10 (though it usually has low autocorrelation length).

\begin{figure}[h!]
    \centering
    \includegraphics[width=0.5\textwidth]{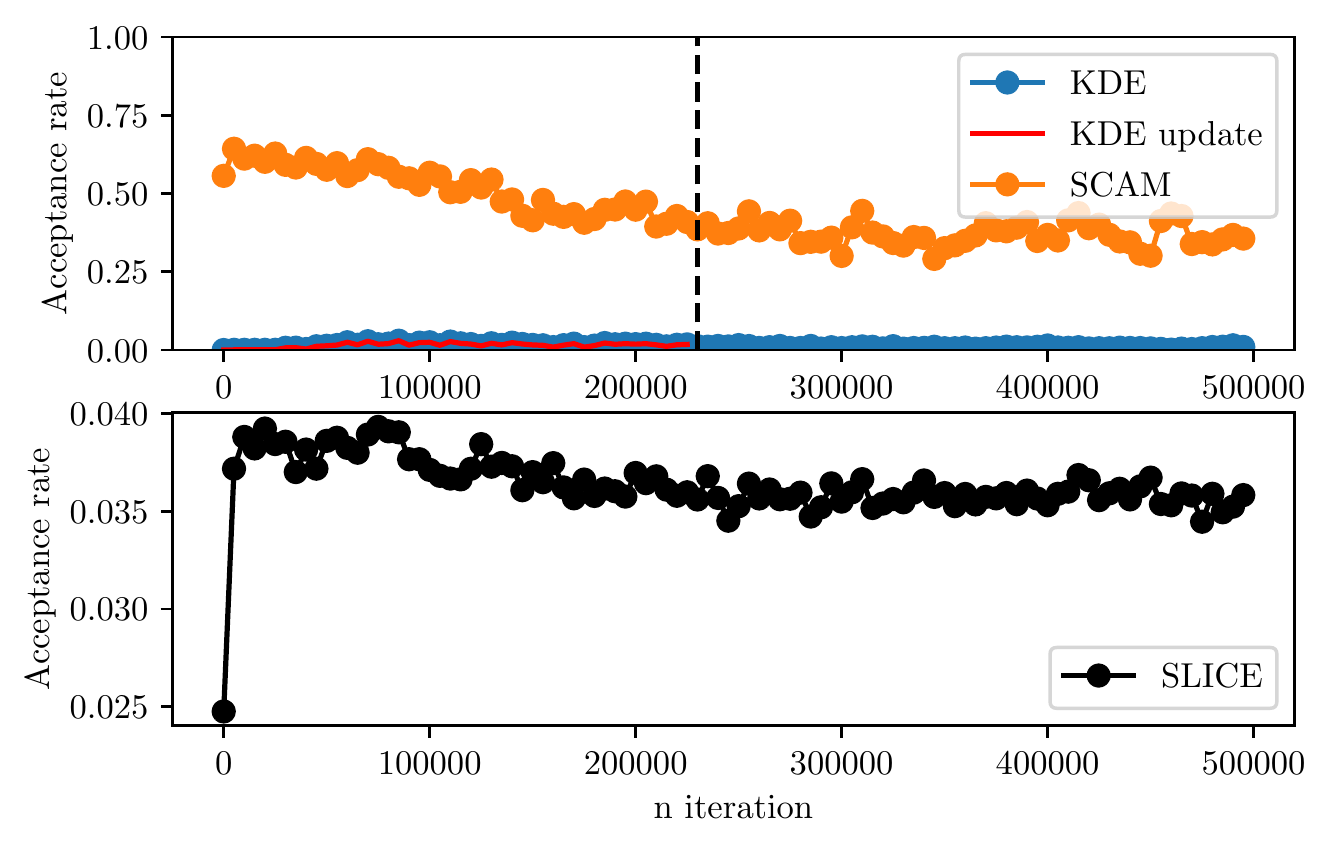}
    \caption{Acceptance rate of all proposals for $n_{kde} = 5$. KDE and SCAM are on top panel, SLICE on bottom panel. Red plot and black dashed line shows where KDE finally converged and stopped updating.}
    \label{fig:ar_nkde_5}
\end{figure}

Figure~\ref{fig:ar_nkde_5} compares acceptance rate for $n_{kde} = 5$. Increase in the dimensionality of the jumps has a drastic effect on the acceptance rate, its value drops to about 0.015.  It is also interesting to compare behaviour of SCAM and slice for two runs. Slice shows very stable/consistent results, while SCAM has significant fluctuations though preserving the trend.  The two-dimensional KDE jumps seem to be the best option in this application. 

\begin{figure}[h!]
    \centering
    \includegraphics[width=0.5\textwidth]{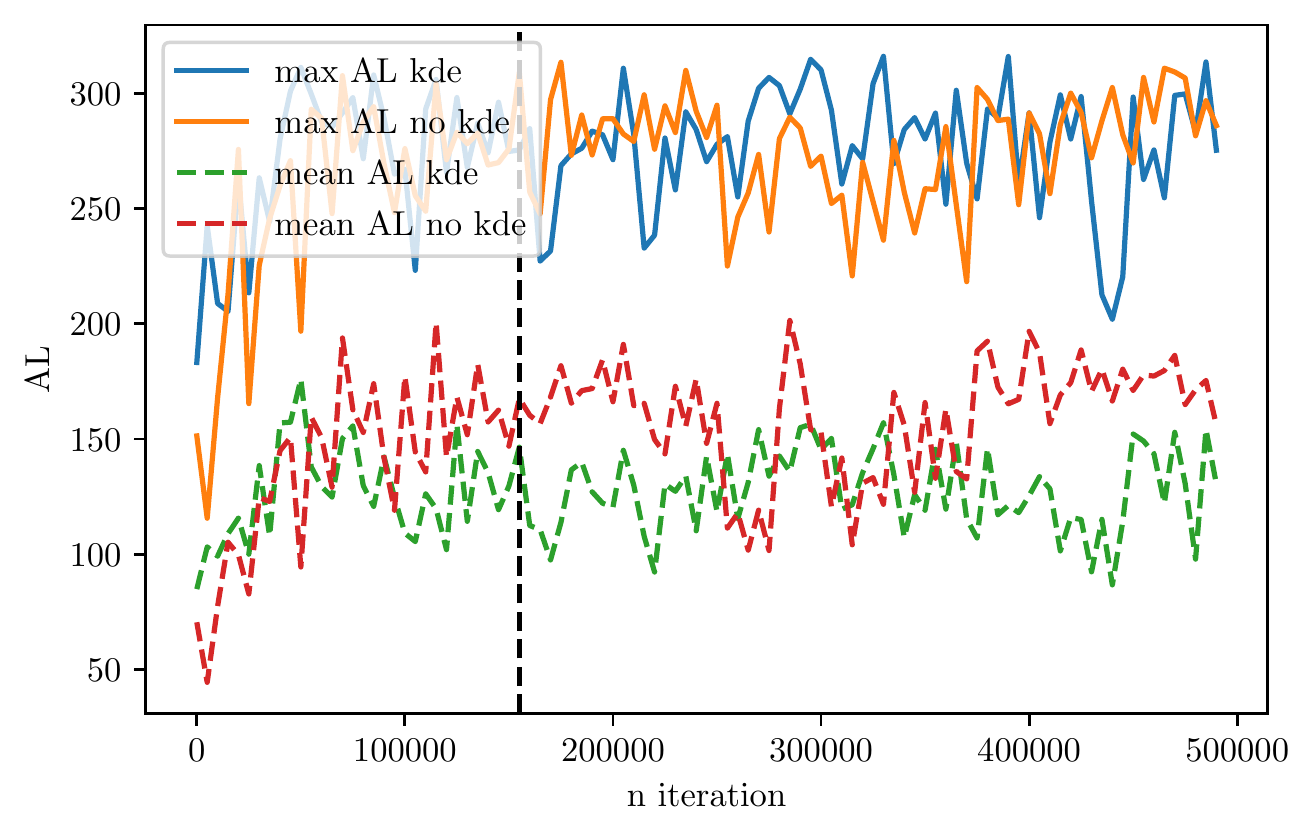}
    \caption{Maximum and mean autocorrelation lengths for $n_{kde} = 1$, comparing a run with KDE against a run without KDE. Black dashed line shows where KDE finally converged and stopped updating.}
    \label{fig:al_nkde_1}
\end{figure}

Next we check the autocorrelation length when we run with and without KDE-based proposal. We restrict ourselves  with the case  $n_{kde} =1$ since  $n_{kde} =5$ has a very poor acceptance rate. 
Figure \ref{fig:al_nkde_1} compares maximum and mean autocorrelation of two runs.   We observe that the mean value is slightly (about 17\%) lower when we include KDE-based proposal  and the maximum length remains the same. As we have already mentioned, mixing SCAM with Metropolis-Hastings and slice steps does reduce the autocorrelation already (compared to PTA example where we did not use slice sampling). In addition we use parallel tempering algorithm, where the hot chains could be seen as yet another jump proposal. All in all, KDE does not add much to already reduced autocorrelation run in the current analysis.  

%\begin{figure}[h!]
%    \centering
%    \includegraphics[width=0.5\textwidth]{figures/al_nkde_5.pdf}
%    \caption{Maximum and mean autocorrelation lengths for $n_{kde} = 5$, comparing a run with KDE against a run without KDE. Black dashed line shows where KDE finally converged and stopped updating.}
%    \label{fig:al_nkde_5}
%\end{figure}
%
%\paragraph{$\boldsymbol{n_{kde} = 5}$}
%
%As expected, because of the very low acceptance rate, the KDE here has little or no effect on the autocorrelation length. The value of $n_{kde}$ seems to be very important and its optimal value depends a lot on the dataset we are working with. A good addition to the method would be to tune this $n_{kde}$ adaptively for optimal performance of the proposal instead of using an arbitrarily fixed $n_{kde}$.

% All in all, does not look like it is going extremely well, slight improvement on the mean AL so not so bad, good acceptance rate, but this case may be too hard, this is as hard as it gets.

% \ref{eq:dkl_converge} is satisfied.

\section{Conclusion}
\label{sec:conclusion}

Bayesian formalism is a usual approach in nowadays gravitational waves data analysis. The inference of the parameters posterior distribution is often done using MCMC and the efficiency strongly depends on the proposal it uses. 
In this article we have presented KDE-based proposal which can be either built on-the-fly during the extended burn-in stage or constructed using posterior points from another run. 

The suggested KDE-based proposal has several extended features: (i) the adaptive bandwith based on the local density of points (small bandwidth in the densely sampled regions of the parameter space); (ii) splitting parameter space into sub-groups of the correlated parameters and applying KDE on each subgroup, we identify correlations using JSD; (iii) possibility of building KDE adaptively.  

We tested this proposal by running MCMC on IPTA data, using a KDE that we have built from previous MCMC runs  (i.e. non adaptive case).  The advantage of KDE-based proposal was clearly seen in high acceptance rate with low autocorrelation length. We have found that using rather low value $n_{kde} = 1-5$ (number of subgroups used in the jump simultaneously) seems to be optimal. 

Another application of the KDE-based proposal was  in running PTMCMC on the simulated LISA data searching for Galactic white dwarf binaries in a narrow frequency band. In this case we have built KDE adaptively during an extended burn-in stage.  The addition of KDE-proposal to the sampling had only a moderate impact: it shows a decent acceptance rate (about 31\%) with only small improvement in the autocorrelation length. Moreover we have shown that low-dimensional jumps are strongly preferred.

Few things could be improved, most notably in the adaptation. The threshold for grouping parameters was chosen somewhat ad hoc, and correlation of some parameters could be close to the threshold. We did observe the fluctuation in choosing the subgroups during the adaptation. One possibility could be to choose not one but two plausible grouping, build KDE for each and use two KDE proposals in a probabilistic manner. The criteria for stopping adaptation 
was also chosen somewhat arbitrary, and might benefit from the further tuning.  Finally, we should implement an adaptive tuning for optimal $n_{kde}$ based on the acceptance rate.

%are yet to be improved for this proposal but we will focus here on two points. First, we need a stronger convergence criterion for our adaptive KDE to be sure that whenever we stop updating, we get the best possible quality proposal (computing KL may not be necessary). Counting how many times a parameter group is found could be enough. We could even build several KDEs from the most probable groupings and choosing one of them according to their odds ratio before performing a MCMC step. It would give a more general representation of the data and most likely recover lost correlations between one group of parameter and another. Finally, we should implement an adaptive $n_{kde}$ method so we optimize the efficiency of our KDE. The latter could be realized by keeping track of the acceptance rate and tuning $n_{kde}$ accordingly so we get as close as possible to the optimal acceptance rate of 0.234 \cite{optimal_ar}.

\bibliographystyle{plain}
\bibliography{biblio.bib}

%\section{Appendix}

\appendix
 
\section{Kullback-Leibler divergence}
\label{A:kl}

We test the quality of our KDE by computing the KL divergence for each subgroup $\hat{f}_{\alpha} (\textbf{x}_{\alpha}, \vec{h}_{\alpha})$. Because we cannot know the true pdf for a set of samples \{$\vec{X}$\}, we have to resort to binning. From sub-KDE $\hat{f}_{\alpha} (\textbf{x}_{\alpha}, \vec{h}_{\alpha})$ built on subgroup of parameter samples \{$\vec{X}_\alpha$\}, we draw a new set of samples \{$\vec{X}^* _\alpha$\}. We estimate the corresponding normalized histogram distributions $\textrm{P}_\alpha$ and $\textrm{P}^* _\alpha$ using same grid of $N$ bins to compute the KL divergence as \cite{kl} :

\begin{equation}
	\textrm{KL}(\textrm{P}_\alpha||\textrm{P}^* _\alpha) = \sum_{i=0} ^N \textrm{P}_{\alpha, i} \bigg( \ln \textrm{P}_{\alpha, i} - \ln \textrm{P}^*_{\alpha, i} \bigg).
\end{equation}
 
To avoid divergence of the logarithms, we set every $\textrm{P}_{\alpha, i}$ and $\textrm{P}^*_{\alpha, i}$ that are equal to 0 to the minimum found value in $\textrm{P}_{\alpha} \bigcup \textrm{P}^*_{\alpha}$ that is not 0.
 
\section{Optimal bandwidth of KDE}
\label{A:optimband}.

We start this appendix with defining few useful expressions that will be used in our derivations later.
\begin{itemize}
	\item The overlap between two neighbouring kernels of same bandwidth is given by
	\begin{equation}
		\boxed{\int d \textbf{x} K(\textbf{x}-\vec{X}_a, \vec{h}) K(\textbf{x}-\vec{X}_b, \vec{h}) = \prod_{i=1}^d \frac{\exp{-\frac{1}{4} \frac{|\vec{X}_a-\vec{X}_b|^2_i}{h^2_i}}}{\sqrt{\pi} 2 h_i}}.
		\label{eqn:overlap_integral}
	\end{equation}
	Let us remind you that $d$ is the dimensionality.
	
	\item If a set of samples $\{\vec{X}\}$ of size $N$ drawn from the probability density function $f(x)$, then we can approximate the averaging integral as :
	
	\begin{equation}
		\boxed{\int dx f(x) g(x) \simeq \frac{1}{N} \sum _a g(X_a)},
	\end{equation}
	where the function $g$ is evaluated at the sample points $\vec{X}_a$.
\end{itemize}

The main objective of this Appendix is to derive the optimal local bandwidth which is defined through the minimization
of the mean square error:

\begin{equation}
	\begin{aligned}
		\epsilon ^2  = &\int dx (\hat{f}(x, h) - f(x))^2 \\
		= &\int dx \hat{f}^2(x, h) - 2\int dx \hat{f}(x, h) f(x) + \int dx f^2(x) \\
		= &\int dx \hat{f}^2(x, h) - 2\int dx \hat{f}(x, h) f(x) + const,
	\end{aligned}
\end{equation}
where  $f(x)$ is the true PDF, $(\hat{f}(x, h)$ is its KDE approximation and $const$ is the term independent of the bandwidth $h$.  We introduce local bandwidth $h_a$ attached to each sample point $X_a$.  Next we assume that 
all points in the vicinity of each point have similar bandwidth, in other words, $h_b \approx h_a$ for $k_{near}$ local points $X_b$.  Using these assumptions  we can approximate the first term: 

\begin{eqnarray}
	N^2 \int dx  \hat{f}(x, h_a)  \hat{f}{x, h_b} = \sum_a \prod_{i=1}^d \frac1{2\sqrt{\pi} h_{a,i}} +  \nonumber \\
	\sum_a \sum_{b\ne a} \prod_{i=1}^{d} 
	\frac{ e^{-\frac1{4}  	\frac{\Delta X^2_{ab, i}} {h^2_{a,i}}   } } 
	{2\sqrt{\pi} h_{a,i}},
\end{eqnarray}
 where $h_{a,i}$ is i-th component of bandwidth attached to Gaussian kernel at $\vec{X}_a$ and $\Delta X_{ab,i}  =
 (\vec{X}_a-\vec{X}_b)_i$ is i-th component of the a vector connecting two samples in the parameter space.
 
 Using now second bullet equation and excluding the actual sample from the sum (for improving stability and removing the possible bias, see "leave one out estimator" \cite{optimal_bw}) we obtain for the second term

\begin{eqnarray}
	2  \int dx  \hat{f}(x, h_a)  \hat{f}{x} \approx \frac2{N^2} 
		\sum_a \sum_{b\ne a} \prod_{i=1}^{d} 
	\frac{ e^{-\frac1{2}  	\frac{\Delta X^2_{ab, i}} {h^2_{a,i}}   } } 
	{\sqrt{2\pi} h_{a,i}},
\end{eqnarray}
where we have assumed $N\gg 1, \;\; N(N-1) \approx N^2$. Combining these terms together gives us 

\begin{eqnarray}
\begin{aligned}
	N^2 (\epsilon^2 - const) = \sum_a \left\{    \prod_{i=1}^d \frac1{2\sqrt{\pi} h_{a,i}} + 
	\sum_{b\ne a} \prod_{i=1}^{d} 
	\frac{ e^{-\frac1{4}  	\frac{\Delta X^2_{ab, i}} {h^2_{a,i}}   } } 
	{2\sqrt{\pi} h_{a,i}} -  \right. \nonumber \\
	\left. 2 \sum_{b\ne a} \prod_{i=1}^{d} 
	\frac{ e^{-\frac1{2}  	\frac{\Delta X^2_{ab, i}} {h^2_{a,i}}   } } 
	{\sqrt{2\pi} h_{a,i}}	
	  \right\}.
\end{aligned}
\end{eqnarray}
Find the minimum of this expression by differentiating with respect to $h_{c,j}$ and equating it to zero: 

\begin{eqnarray}
\begin{aligned}
0 =  -\frac1{h_{c,j}} \prod_{i=1}^d \frac1{2\sqrt{\pi} h_{c,i}} \left\{ 1 +
\sum_{b\ne c} \left[
1 - \frac1{2} \frac{\Delta X^2_{bc, j}}{h^2_{c,j}}
\right]  
\prod_{i=1}^{d} 
 e^{-\frac1{4}  	\frac{\Delta X^2_{bc, i}} {h^2_{c,i}}   } %}
  \right. \nonumber \\
\left. 
-2(\sqrt{2})^d  \left[
1 -  \frac{\Delta X^2_{bc, j}}{h^2_{c,j}}
\right]  
\prod_{i=1}^{d} 
e^{-\frac1{2}  	\frac{\Delta X^2_{bc, i}} {h^2_{c,i}}   } 
\right\}
\end{aligned}
\end{eqnarray}

Next we assume quite conservative approximation:  $\frac{\Delta X^2_{bc, j}}{h^2_{c,j}} \ll 1$ for all points $\vec{X}_b$ in vicinity of $\vec{X}_c$ and all components $j$.  This assumption overestimates the bandwidth and therefore conservative: this is what is used in this paper.  Expanding in this small parameters and retain only the terms quadratic in this small ratio we obtain 
the system of linear equations for $1/h^2_{c,j}$:

\begin{eqnarray}
	\frac{k_{near}( 2^{d/2+1} -1) -1} {(2^{d/2} - 1 ) } = \sum_{b\ne c} \left[
	3\frac{\Delta X^2_{bc, j}}{h^2_{c,j}}  + \sum_{i\ne j} \frac{\Delta X^2_{bc, i}}{h^2_{c,i}}
	\right]	.
\label{eq:opt_bw_lin_system}
\end{eqnarray}
 Solving this system at each point $\vec{X}_c$ for each direction in the parameter space ($j$) gives us the desired local  bandwidth $\vec{h}_c$.
 
As an alternative approach we can assume that the bandwidth is comparable to the distance to the neighnbours and
 define $h^2_{c,j} = \overline{\Delta X^2}_{c,j}(1 + \varepsilon_{c,j})$, where $\overline{\Delta X^2}_{c,j} = 1/k_{near} \sum_b \Delta X^2_{bc, j} $ is the average square distance ($i$-th component)  to the 
points in vicinity of $\vec{X}_c$ and assume that $\Delta X^2_{bc, j}/\overline{\Delta X^2}_{c,j} \sim 1 $  and   $\varepsilon_{c,j} \ll 1$ for all $b,c,j$. This yields

\begin{eqnarray}
	\prod_{i=1}^{d} 
	e^{-\frac1{4}  	\frac{\Delta X^2_{bc, i}} {h^2_{c,i}} }  \approx  \left(
	1 + \frac1{4} \sum_{i}^d 	\frac{\Delta X^2_{bc, i}} { \overline{\Delta X^2}_{c,i} }
	\right)
	\prod_{i=1}^{d} 
	e^{-\frac1{4}  	\frac{\Delta X^2_{bc, i}} { \overline{\Delta X^2}_{c,i} }  }.
\end{eqnarray}
Using this approximation we arrive at the system of linear equations for $1/h^2_{c,j}$:
\begin{eqnarray}
\begin{aligned}
-1  + \sum_{b\ne c}  
2(\sqrt{2})^d  
\prod_{i=1}^{d}  e^{-\frac1{2}  	\frac{\Delta X^2_{bc, i}} { \overline{\Delta X^2}_{c,i} }  }  - P_{bc} 
\left(
1 - \frac1{2} \sum_{i}^d 	\frac{\Delta X^2_{bc, i}} { \overline{\Delta X^2}_{c,i} }
\right) = \nonumber \\
\sum_{b\ne c}  P_{bc}\left(
3 	\frac{\Delta X^2_{bc, j}} { \overline{\Delta X^2}_{c,j} } \varepsilon_{c,j} + 
 \sum_{i\ne j}  	\frac{\Delta X^2_{bc, i}} { \overline{\Delta X^2}_{c,i} } \varepsilon_{c,i}
\right),
\end{aligned}
\end{eqnarray}
where 
$$
P_{bc} \equiv \frac1{4}\left[
	\prod_{i=1}^{d} 
e^{-\frac1{4}  	\frac{\Delta X^2_{bc, i}} { \overline{\Delta X^2}_{c,i} }  }
- 4(\sqrt{2})^d
	\prod_{i=1}^{d} 
e^{-\frac1{2}  	\frac{\Delta X^2_{bc, i}} { \overline{\Delta X^2}_{c,i} }  }
\right] .
$$
 Solving this system at each point $\vec{X}_c$ for each direction in the parameter space ($j$) gives us the desired local  bandwidth $\vec{h}_c$.

\newpage

%\stb{BELOW is old text left there for comparison}
%
%
%
%
%%\subsection{Optimal bandwidth approximation}
%
%We will search for the optimal bandwidth by minimizing the MSE given as 
%%, as explained in section \ref{sec:opt_bw}. Using the expression in equation \ref{eq:mse}, we have :
%
%
%\mf{Do we need to keep this bit of explanation to justify the use of global bandwidth ?}
%The latter applies for a global bandwidth (same for all kernels). We can modify this expression by adding a local variation to each kernel to define a local bandwidth :
%
%\begin{equation}
%    h_i = h +\delta h_i
%\end{equation}
%
%with $<h_i>=h$, the mean bandwidth. In that framework, $h$ identifies as the global bandwidth and still stands as the zero order solution because :
%
%\begin{equation}
%    \epsilon ^2 (h + \delta h_i) = \epsilon^2 (h) + \delta h_i \frac{\partial}{\partial \delta h_i} \epsilon ^2 (h) + ...
%\end{equation}
%
%The new expression for our MSE becomes :
%
%\begin{equation}
%\begin{aligned}
%    \epsilon ^2  = &\int dx (\hat{f}(x, h_i) - f(x))^2 \\
%     = &\int dx \hat{f}(x, h_i)\hat{f}(x, h_j) - 2\int dx \hat{f}(x, h_i) f(x) + K
%\end{aligned}
%\label{eq:mse_local}
%\end{equation}
%
%Solving this linear system for each point in the dataset gives the local bandwidth $h_i$ of the kernel centered on point $\vec{X}_i$.
%
%If one prefers to work with a global bandwidth, it is still possible to obtain it by taking the average of all local bandwidths.

\section{The M3C2 sampler}
\label{sec:appendix_sampler}

The \verb|M3C2| \footnote{\url{https://gitlab.in2p3.fr/lisa-apc/mc3}}
(Multiple parallel Markov Chain Monte Carlo) is a python implementation of
MCMC sampler.  The aim of this tool is to improve the sampling
robustness of complex posterior distribution, by running multiple
chains in parallel. The cross check of the chain performance informs
us about convergence (using Gelman-Rubin ratio \cite{10.1214/ss/1177011136}).  We
have implemented two mechanisms of building the chain (1) using slice
sampling (\verb|slice| ) and (2) Metropolis-Hastings algorithm (MH)
which could be used separately or together improving the mixture of
the chains and reducing the auto-correlation length.  Even though the
sampler is very generic, we primarily use it within the context of GW
data analysis. For the Metropolis-Hastings method we have implemented
a set of proposal jumps:
\begin{itemize}
\item \verb|SCAM| (Single Component Adaptive Metropolis), jumps along
  one randomly chosen direction given by the eigen vectors of the
  covariance matrix \cite{roberts_rosenthal, SCAM}
\item \verb|DE| (Differential Evolution), jumps along the direction
  given by difference of two randomly chosen samples of the chains, 
  or (as in the classic implementation \cite{Braak, braak_vrugt}) by using state of different chains running in parallel.
\item \verb|ReMHA| (Regional Metropolis Hastings Algorithm), the
  proposal represented by a mixture of several Gaussians
  distributions~\cite{roberts_rosenthal, DOI:10.1198/jasa.2009.tm083}.
\end{itemize}

For \verb|SCAM| we build the covariance matrix adaptively based on the
accumulated samples. The use of the accumulated samples breaks the
Markov property of the chain, making it asymptotically Markovian. The
stability of the covariance matrix is yet another sign of the
converged chain. One can stop adaptation after burn-in run.  This
proposal is suggested in \cite{Ellis}.

\verb|ReMHA| is similarly used adaptively. We use accumulated samples
during the burn-in to estimate the number of clusters using
Variational Bayesian Gaussian Mixture (\verb|skikit-learn| package)
and use this Gaussian mixture probablity as a proposal. This proposal
is somewhat similar to the one suggested in
\cite{DOI:10.1198/jasa.2009.tm083}.
 
\verb|DE| could be used as a proposal (snooker) described in \cite{braak_vrugt} using multiple chains running in parallel or using the accumulated samples for each chain to propose the jump.
 There is no much difference between those two ways in case of well
converged chains. However, the behavior and efficiency of those two implementations 
is very during the burn-in stage.
 
In \verb|slice| sampling, we use slicing of the parameter space either
randomly or along the eigen directions of the covariance matrix, the
choice is made with a probability set by user. In case of mixture of
\verb|slice| and MH, the frequency of each method is defined by a user
specified weight. In addition to preset proposals available in
\verb|M3C2|, user can add custom jump-proposals using a common
interface.  The weights and proposals can be set individually for each
running chain.

Besides running parallel independent chains, \verb|M3C2| sampler has
also parallel tempering implementation with an adaptive temperature
ladder following \cite{2016MNRAS.455.1919V}. The adaptation is aiming
at increasing the acceptance rate between the chains.
 
The multi-chain scheme of \verb|M3C2| can be easily deployed on the
multicore CPU infrastructure.  Data exchange between chains, in case
of parallel tempering, is restricted to its minimum level (pairwise
communication between the chains), to ensure a good scalability.

\end{document}